\documentclass[12pt,preprint]{aastex}

\def\feh{{\rm [Fe/H]}}

\def\Teff{T_{\rm eff}}

\def\popi{Population~I}
\def\popii{Population~II}

\def\Str{Str\"omgren}

\begin{document}

\title{The \emph{uBVI} Photometric System. II. Standard Stars}

\author{Michael H.\ Siegel\altaffilmark{1} and Howard E.\ Bond\altaffilmark{2}}

\affil{Space Telescope Science Institute, 3700 San
Martin Drive, Baltimore, MD 21218; msiegel@stsci.edu, bond@stsci.edu}

\altaffiltext{1}
{Current address: University of Texas - McDonald Observatory,
Austin, TX 78712; e-mail: siegel@astro.as.utexas.edu}

\altaffiltext{2}
{Visiting Astronomer, Kitt Peak National Observatory and Cerro Tololo
Interamerican Observatory, National Optical Astronomy Observatory, which are
operated by the Association of Universities for Research in Astronomy, Inc.,
under cooperative agreement with the National Science Foundation.}

\begin{abstract}
Paper I of this series described the design of a CCD-based photometric system
that is optimized for ground-based measurements of the size of the Balmer
discontinuity in stellar spectra. This ``{\it uBVI\/}'' system combines the
Thuan-Gunn $u$ filter with the standard Johnson-Kron-Cousins \textit{BVI} filters, and
it can be used to discover luminous yellow supergiants in extragalactic systems
and post-asymptotic-giant-branch stars in globular clusters and galactic halos.
In the present paper we use \textit{uBVI} observations obtained on 54 nights with 0.9-m
telescopes at Kitt Peak and Cerro Tololo to construct a catalog of standardized
$u$ magnitudes for standard stars taken from the 1992 catalog of Landolt.  We
describe the selection of our 14 Landolt fields, and give details of the
photometric reductions, including red-leak and extinction corrections,
transformation of all of the observations onto a common magnitude system, and
establishment of the photometric zero point.  We present a catalog of $u$
magnitudes of 103 stars suitable for use as standards. We show that data
obtained with other telescopes can be transformed to our standard system with
better than 1\% accuracy.
\end{abstract}


\section{Introduction}

Paper~I in this series (Bond 2005) described the design of a new ground-based
CCD photometric system which is highly optimized for measurement of the Balmer
jump in stellar energy distributions.  This ``\textit{uBVI}'' system combines the $u$
bandpass of Thuan \& Gunn (1976) with the standard broad-band
Johnson-Kron-Cousins $B$, $V$, and $I$ filters.  As shown in Paper~I, the
Thuan-Gunn $u$ filter has a higher figure of merit for measuring the Balmer 
jump than the other commonly used ultraviolet filters (\Str\ $u$, the Sloan
Digital Sky Survey $u'$, and Johnson $U$), because of its high throughput
combined with little transmission longward of the Balmer discontinuity. The
Johnson-Kron-Cousins filters were chosen at longer wavelengths because of the
vast legacy of \textit{BVI} stellar photometry, the broad bandpasses and high
throughputs, and the extensive network of well-calibrated standard stars
established through the work of Landolt and others.

The scientific motivation for a highly efficient filter system that measures
the Balmer jump was presented in Paper~I.  To summarize, the most
luminous stars in both young and old populations are objects of
low surface gravities that have large Balmer discontinuities. In \popi\
many of the visually brightest non-transient stars are yellow supergiants of
spectral types A, F, and early G, which can attain absolute magnitudes of
$M_V\simeq-10$ (Humphreys 1983). The brightest stars of \popii\ are
post-asymptotic-giant-branch (PAGB) stars that have left the tip of the AGB and
are evolving through spectral types F and A at absolute magnitudes of
$M_V\simeq-3.3$ (Bond 1997; Alves, Bond, \& Onken 2001).  Both classes of stars
show considerable promise as tracers of their respective populations and as
extragalactic standard candles. 

Paper~I presented a calibration of \textit{uBVI} photometry against the basic stellar
parameters of effective temperature, surface gravity, and metallicity, based on
theoretical model stellar atmospheres.  It was shown in that paper that the
$u-B$ color index is very sensitive to $\log g$ for stars with
$\Teff\simeq5$,000 to 10,000~K\null. 

Proper exploitation of the system, however, requires a set of standard stars for
the calibration of \textit{uBVI} photometry. For $B$, $V$, and $I$, there is, of
course, the widely used catalog of faint equatorial standards produced by
Landolt (1992, hereafter L92).  

Unfortunately, the Thuan-Gunn $u$ band (hereafter referred to simply as ``$u$'')
did not share in this happy state of affairs at the beginning of the work
described here.   Thuan \& Gunn (1976) provided a listing of $u$ magnitudes in
their system for over two dozen standard stars, but these stars are extremely
bright for CCD work ($V\simeq8$--11), many of them are inaccessible from the
southern hemisphere, and in general they are not as well calibrated in \textit{BVI} as
the L92 equatorial standards. To our knowledge, the initial photometric
standards of Thuan \& Gunn have only been augmented by Kent (1985), but again
these stars are bright and many are accessible only from the north.  J\o rgensen
(1994) established secondary standard stars for the Thuan-Gunn system, but
omitted the $u$ band.  More recently, the SDSS ultraviolet filter ($u'$) has
been extensively calibrated in the course of the SDSS survey, but of course it
is a different filter, with significant transmission above the Balmer jump and
hence a lower figure of merit for measuring the size of the discontinuity (see
Paper~I for details). It was thus apparent that we would have to establish our
own new and independent system of faint standard stars in the $u$ band for
calibration of our CCD-based \textit{uBVI} photometry.

Since we were using the standard \textit{BVI} filters, we made the obvious decision to
use standard stars from L92 to calibrate those three bandpasses.  We therefore
decided  to base the $u$ calibration on the same set of standard stars.

This paper describes the calibration process, and presents the list of standard $u$ 
magnitudes for the L92 stars that resulted.  \S2 of this paper describes the 
observations and data reduction; \S3 discusses the calibration of the measures to 
the \textit{BVI} system; \S4 describes the process of creating the catalog of 
standard $u$ magnitudes; \S5 presents the catalog of $u$ magnitudes for the 
standard stars; \S6 gives an example of the transformation of \textit{uBVI} 
photometry to our standard system; and \S7 summarizes.

\section{Standard-Star Observations and CCD Reductions}

Our observational data consist of CCD \textit{uBVI} frames obtained with the
0.9-, 1.5-, and 4-m telescopes at Cerro Tololo Interamerican Observatory
(CTIO), and the \hbox{0.9-}, 2.1-, and 4-m telescopes at Kitt Peak National
Observatory (KPNO), during a series of observing runs from 1994 through 2001.
The primary aim of these runs was a large-scale search for PAGB stars in the
halos of nearby galaxies and in Galactic globular clusters.

For purposes of calibration, Landolt fields selected from those listed in
Table~1\footnote{Finding charts, coordinates, and \textit{UBVRI} photometry  for these
stars can be found in L92.} were observed on every photometric night on which
our program fields were observed. The fields listed in Table~1 were specifically
chosen to contain several stars that lie within a few arcminutes of each other
and are thus observable with a single CCD pointing.  They also included at least
one blue star, and the fields are distributed approximately uniformly around the
celestial equator.  We observed these fields intensively over an interval of
seven years, and the $u$ magnitudes have been reduced to a standard system as
described in detail in this paper.  The list of standard stars in these fields,
presented below, now forms the basis for calibration of $u$-band photometry from
any telescope in either hemisphere.

The majority of our data was taken with the 0.9-m telescopes at CTIO
and KPNO, and therefore we will base the calibration of the \textit{uBVI} system on
the observing runs from these two telescopes.

On the CTIO 0.9-m, we used the $2048\times2048$ ``Tek3'' CCD.  On the KPNO 0.9-m
we used the $2048\times2048$ ``T2KA'' CCD\null. The fields of view for these two
chips are $13'\times13'$ for Tek3 at CTIO, and $23'\times23'$ for T2KA at KPNO;
the plate scales are $0\farcs396 \,\rm pixel^{-1}$ and $0\farcs688 \,\rm
pixel^{-1}$, respectively. All of the $u$-band frames on both telescopes were
obtained with a custom-built $4\times4$-inch filter, described in detail in
Paper~I\null.  For \textit{BVI}, we used filter sets provided by KPNO and CTIO, with
the following filter identifications: at CTIO: B-Tek2, V-Tek2, and I-KC31; at
KPNO: 1569, 1421, 1444.

Table~2 gives details of the 15 0.9-m observing runs used for standardizing the
\textit{uBVI} system. Columns 3 through 6 give the number of nights on which
photometric observations of standard stars were obtained, the number of $u$-band
CCD standard-star frames obtained during the run, the number of distinct
standard stars observed during the run, and finally the total number of
individual $u$ magnitudes measured.  Standard-star observations were made on 54
completely or partially photometric nights, on which 1075 CCD frames (271 of
them in the $u$ band) of Landolt fields were obtained, providing 1738 $u$-band
measures of 142 potential standard stars.

We began the reductions by bias-subtracting, trimming, and flat-fielding all of
the data frames using the standard IRAF\footnote{IRAF is distributed by the
National Optical Astronomy Observatory, which is operated by the Association
of Universities for Research in Astronomy, Inc., under cooperative agreement
with the National Science Foundation.} procedures in the {\it ccdproc\/} and
{\it quadproc\/} packages.  As noted in Paper~I, the flat-fielding in the $u$
band must be done using frames exposed on the twilit evening and morning sky;
for $B$, $V$, and $I$ we used both twilight and dome flats.

\section{\emph{BVI} Calibration}

The first step in deriving the $u$ standard system was the transformation of the
\textit{BVI} instrumental measures to the L92 system.  Apart from being necessary
to derive calibrated photometry for our program stars, this step allowed an
independent check on the photometric quality of the observing nights.

Transformation of the \textit{BVI} instrumental magnitudes began with the interactive
identification of the L92 standard stars in each CCD field.  We used the DAOPHOT
package (Stetson 1987) to measure aperture photometry on each star for a series
of apertures over a range of radii up to $14''$ for the CTIO data and $24''$ for
the KPNO data (because of the larger pixel scale).\footnote{For the $u$
magnitudes discussed below, we found that magnitudes were typically only usable
out to radii of $\sim$$5''$ because of the smaller stellar signals.} We did not
subtract off neighboring stars before performing the multi-aperture photometry,
because the standard system of L92, based on photoelectric aperture photometry,
includes the light of any neighbors lying within the apertures.  In practice,
this choice has little noticeable effect since most of the chosen L92 standards
lack nearby companions.

We used DAOGROW (Stetson 1990) to perform curve-of-growth analyses for each
observing run.  We did not, however, use aperture corrections for the standard
stars, but instead used the total extrapolated magnitude measures produced by 
DAOGROW\null.  We have found that these extrapolated magnitudes provide the best
measures for standard-star calibration, since they measure nearly all of the
light from each star, independently of small changes in seeing and focus.  We
have found that this method noticeably reduces the scatter in the photometric
measurements.

In agreement with other authors (L92; Johnson \& Bolte 1998) we found that two
L92 ``standards,'' PG~1047+003C and PG~1323$-$086A, are low-amplitude variables,
and we eliminated them at this stage. We also note that PG~1047+003 itself has
been found to be a rapidly oscillating variable star (O'Donoghue et al.\ 1998),
but given its small amplitude and short periods it remains usable as a
photometric standard. 

The final \textit{BVI} photometric transformation equations were derived using the
iterative technique described in Siegel et~al.\ (2002).  In this procedure, we
do a simultaneous, interactive fit to the extinction, color, and zero-point
terms via matrix inversion, through a code kindly provided  by C.~Palma. 
Airmass values were derived for each frame using the IRAF procedure {\it 
setairmass\/}, which calculates the photon-weighted airmass. We determined a
single extinction coefficient in each filter for each observing run, and
compensated for residual nightly variations in extinction by allowing the zero
points to vary from night to night.  Fixing the zero points and allowing the
extinction coefficients to vary from night to night, which would have allowed
for nightly variations, produced less  internally consistent transformations and
resulted in significant differences  between photometry obtained on different
nights.

In general, we were able to fit the L92 standard system to a precision of better
than 0.01~mag. A handful of nights had slightly poorer transformations (but
still with uncertainties $\leq$0.02~mag) because of slightly poorer observing
conditions or a small number of observations.  We found no need for non-linear
extinction or color coefficients in $B$, $V$, and $I$ during the calibration
process.  

\section{Creation of the \emph{u}-Band Standard Magnitudes}

Having reduced the \textit{BVI} photometry for all of our standard-star observations
and thereby having restricted ourselves to data obtained on demonstrably
photometric nights, we turned to the establishment of a system of standard $u$
magnitudes.

There are four steps in this process: (1)~correction for the red leak in the
$u$ filter, (2)~correction for atmospheric extinction, (3)~unification of the
outside-atmosphere $u$ magnitudes from all of the observing runs on both
telescopes onto a single consistent scale, and (4)~adoption of a zero-point
correction.  Each of these steps is described in detail below.

\subsection{Red Leak}

As discussed in Paper~I, the $u$ filter has a small red leak at about
7100~\AA\null. In the interest of maximizing throughput in the main ultraviolet
band, and for maximum simplicity and economy, we made no attempt to block the
red leak. Instead, we modeled it using the technical specifications of the
cameras, filters, and detectors and model stellar atmospheres, as described in
detail in Paper~I\null. Since the red leak lies in the blue wing of the $I$
bandpass, it is simplest to model the leak as a fraction of the $I$-band
signal, with a small color term. The following corrections were calculated:
\begin{equation}
{\rm RL}/I = 2.27 \times 10^{-4} - 6.80 \times 10^{-5} (B-V) + 5.36
\times 10^{-5}  (B-V)^2 -3.00 \times 10^{-5}  (B-V)^3
\end{equation}
\begin{equation}
{\rm RL}/I = 2.00 \times 10^{-4} - 6.26 \times 10^{-5}  (B-V) + 5.23
\times 10^{-5}  (B-V)^2 - 2.89 \times 10^{-5}  (B-V)^3
\end{equation}
for the CTIO and KPNO 0.9-m telescopes, respectively. Here RL is the number of
$u$ electron counts from the red leak, 
$I$ is the electron count measured in the $I$ band, and
$B-V$ is the color of the star on
the standard system.
All photon counts are per unit time and are measured inside the atmosphere. 
Therefore, the first step in the reductions was to subtract RL from the
inside-atmosphere $u$ electron counts, using either eq.~1 or~2.

The red-leak correction is of little practical importance in most applications,
since the \textit{uBVI} system will usually be used for fairly blue stars; thus this
step can usually be omitted.  Nevertheless, we have included this correction in
the reduction of the standard stars in order to improve the realism of the
photometric system. As noted in Paper~I, the red leak contributes about 1\% of
the total $u$ signal for stars with $B-V=0.8$, and rises to 10\% of the signal
at $B-V=1.47$. We have eliminated from our standard-star catalog all stars with
L92 $B-V$ colors redder than 1.45. 

Our $u$-band standard-star observations usually had matching $I$-band
observations, so that eqs.~1 or 2 could be applied. In a few cases, there was
either no $I$ image obtained at the same time, or the corresponding $I$ stellar
image was saturated; these measures were discarded.

\subsection{Atmospheric Extinction}

The effects of atmospheric extinction were modeled in Paper~I, where it was
shown that because of the steep dependence of the extinction on wavelength
across the $u$ band, it is desirable to include both a small color term and a
small non-linear airmass term in the extinction equation. The equation is thus  
\begin{equation} 
u_{\rm instr}(X) = u_{\rm out} + [a + b(B-V)] X + k_2 X^2 \, , 
\end{equation} 
where
$u_{\rm instr}(X)$ is the instrumental $u$ magnitude measured at airmass $X$,
$u_{\rm out}$ is the $u$ magnitude outside the atmosphere, $a + b(B-V)$ and
$k_2$ are the linear and quadratic extinction coefficients, and $B-V$ is the
stellar color on the standard system.  In principle, all three coefficients
could be fit directly if enough stars of a wide range of color were observed
over a broad range of airmass.

In practice, we initially fixed $b$ and $k_2$ to the theoretical values derived
in the simulations of Paper I ($-0.033$ and $-0.007$, respectively) and
determined $a$ directly for each observing run by intercomparing stars observed
over a range of airmass.  As noted in \S3, we adopted an average $a$ coefficient
for each observing run; these ranged from 0.550 to 0.620 mag~airmass$^{-1}$ over
the duration of this project.  The $u$ extinction coefficients were found to be
well-correlated with the extinction in the \textit{BVI} filters.

After the initial solution for $a$, we intercompared the nights of each
observing run to identify any night-to-night changes in photometric zero point. 
These changes were generally small (usually a few hundredths of a magnitude, but
occasionally as large as 0.15~mag), and consistent with the previously derived
zero-point shifts in the \textit{BVI} calibration.   We applied these zero-point
corrections, then derived the extinction coefficients again, this time allowing
both $a$ and $b$ to vary.  The $a$ terms were little changed from the initial
solution, but there was a marked decreased in the scatter.  The average $b$ term
was $-0.0357$, in excellent agreement with the simulation.  We  iterated between
nightly zero-point corrections and extinction corrections until all terms
converged to 0.001~mag. We left $k_2$ fixed at the theoretical value since
generally there were not enough observations of standard stars over a wide
enough variety of airmasses for an empirical determination.

\subsection{Unifying the Instrumental $u$-Band Photometry}

The next step is to combine all of the $u$ magnitudes from all of the observing
runs into a catalog of standard values. It would be insufficient simply to
average the instrumental outside-atmosphere $u_{\rm out}$ measurements over all
of the different observing runs, for two reasons. (1)~The CTIO and KPNO
instrumental magnitudes were obtained with two different telescopes, cameras,
and detectors, over intervals of several years; because of differences in the
response functions, the instrumental $u$ magnitudes from the two telescopes
differ systematically (even though the same $u$ filter was used at both
telescopes).  (2)~Moreover, we empirically identified significant zero-point
changes and small color shifts between data obtained during separate observing
runs even with the same telescope, in both $u$ and \textit{BVI}\null. The latter
systematic differences are likely the result of slight, possibly
wavelength-dependent, variations over time in the throughputs of the telescope
systems, and possibly long-term changes in atmospheric extinction and detector
characteristics (including gain), or other more subtle effects.

We began the unification process by identifying the two observing runs that
provided the best mutual agreement in $u$ magnitudes. A statistical weight, $W$,
was defined for each combination of two observing runs as $W = \sum
1/\sigma_i^{2}$, where $\sigma_i$ is the standard deviation in the photometric
measures of the $i$th star that the two runs have in common.  The two runs whose
comparison produced the greatest statistical weight were then combined.

The combination was done as follows. First, we calculated a transformation (zero
point and color term) from one run to the other, using a least-squares fit of
the form 
\begin{equation}
u_{\rm 1} = u_{\rm 2} + c + d (B-V) \, ,
\end{equation}
where $u_{\rm 1}$ and $u_{\rm 2}$ are the error-weighted average
magnitudes in observing runs~1 and~2, respectively. This transformation was then
applied in a weighted fashion to all of the magnitude measures in both observing
runs as follows:
\begin{equation}
u_{\rm 1}' = u_{\rm 1} - [W_2/(W_1+W_2)] \, [c + d (B-V)] \, ,
\end{equation}
\begin{equation}
u_{\rm 2}' = u_{\rm 2} + [W_1/(W_1+W_2)] \, [c + d (B-V)] \, ,
\end{equation}
where the statistical weights of the individual observing runs were calculated
from the scatter within each run using $W_1 = \sum 1/\sigma_{i,1}^2$ and $W_2 =
\sum 1/\sigma_{i,2}^2$.  These transformations placed both observing runs onto
the same instrumental system, which is intermediate between the systems of the
two runs.  We then combined all of the measures to produce a new ``observing
run,'' containing a weighted average of all of the stars in common, along with
the transformed measures for the stars not in common.

This new ``observing run'' was then placed back into the set of observing runs,
and the next-best pair of runs was identified. The process described above was
repeated until all 15 runs had been incorporated.

The resulting catalog was then refined by re-determining the coefficients $c$
and $d$ that transform each of the 15 individual runs to the catalog of mean $u$
magnitudes, and applying that transformation to all the individual magnitude
measures.  The purpose of this step was to put all of the independent measures
on the same instrumental system in order to identify interactively any 
discrepant individual measurements (e.g., due to cosmic-ray hits, short-term
atmospheric transparency variations, or other statistical fluctuations).  These
discrepant measures could only be recognized after an initial global solution
had been made.  We rejected 41 individual outliers. 

We finally re-iterated the run-to-run solution by deriving a matrix-inversion
fit of each run to the mean catalog, using the full transformation
equation\footnote{Eq.~5 in Paper~I also included a quadratic color term,
$e(B-V)^2$, but this is necessary only when the measurements have not been
corrected for red leak; see \S6 below.} from Paper~I,

\begin{equation}
u_{\rm instr} = u_{\rm std} + [a + b (B-V)] X + k_2 X^2 + c + d (B-V) \, .
\end{equation}

The transformation color term, $d$, was now small, its largest departure from
zero among all the runs being only $-0.023$.  The final magnitude measures for
each star are the weighted averages of the transformed magnitudes.

As a check on the pipeline, we ran the $B$-band data through the same steps to
determine if we could recover the L92 values.  During the derivation, we
reproduced measures of extinction, zero-point, and color terms that were close
to those derived by simple matrix inversion in \S4.  When these terms were
applied, the system was internally consistent to well within 1\%.  External
comparison to the L92 values was worse but still within 1\%.  The slightly
poorer external comparison is likely the result of our $B$ filter being a poor
match to Landolt's, which produced strong color terms in the transformation (and
opposite signs for KPNO and CTIO)\null.  When only CTIO data are considered, the
external comparison is almost as good as the internal.  The result shows that
our pipeline produced an internally and externally consistent instrumental
photometric system.

\subsection{The Zero-Point Correction}

The final step is to apply a zero-point correction to the catalog of mean
outside-atmosphere $u$ magnitudes of the standard stars.  As discussed in
Paper~I, we will follow the precept adopted for the \textit{uvby} system by \Str\
(1963), who set the $u-b$ color of Vega and other A0~V stars to 1.0. Vega itself
is, however, far too bright to observe with our equipment, and only a few of
the L92 standards have colors in the vicinity of $B-V\simeq0.0$.  We therefore
had to adopt an indirect approach to the zero point.

We observed three lightly reddened open clusters of near-solar metallicity: M34,
M41, and M67.  M34 was observed on 1997 September 20 with the KPNO 0.9-m
telescope, M41 on 1995 October 16 with the CTIO 1.5-m, and M67 on 4 nights in
1998--99 with the KPNO 0.9-m. The $u$ magnitudes of the cluster members were
reduced to the instrumental system defined above, and the \textit{BVI} magnitudes to
the standard L92 system.  The color indices were then adjusted for reddening,
using the formulae in \S4 of Paper~I\null. Table~3 lists the three clusters and
the adopted reddenings. 

We then adjusted the zero point of the $u$ magnitude system so that the
dereddened $u-B$ colors of the main sequences of the three clusters would lie,
on average, on the zero-age main-sequence (ZAMS) relation derived in \S3.3 of
Paper~I from theoretical model atmospheres.  We also gave some weight to fitting
the color difference $(u-B)-(B-V)$ (the analog of the \Str\ $c_1$ index) to the
theoretical ZAMS values (see Paper~I), since the color difference has the
advantage of placing the zero-point and reddening vectors nearly orthogonal to
each other. (We also tried an experiment in which we fit {\it both\/} the color
difference and the reddening, and successfully reproduced the $E(B-V)$ values of
Table~3 to within 0.01~mag.)

Figure~1 shows the dereddened measurements for the three clusters, after
application of the zero-point shift, superposed on the grid of theoretical
colors for solar-metallicity stars calculated in Paper~I\null. The overall
agreement with the predicted shape of the ZAMS relation is good, although the
scatter is somewhat higher for the redder stars (which are faint main-sequence
stars with relatively large photometric errors in our short exposures; also,
although we have removed obvious field stars, some field-star contamination
remains).

M67 shows two characteristics worth noting in Figure~1. (1)~Its main-sequence
stars are slightly above the ZAMS relation, especially at $B-V\approx0.5$--0.6.
These stars have already departed from the ZAMS because of the high age of M67,
and are thus expected to have lower gravities than the ZAMS\null. (2)~The bluer
M67 stars are blue stragglers (ignored in adjusting the zero point for our $u$
magnitudes).  Figure~1 confirms that they have surface gravities similar to
those of main-sequence stars, as was first demonstrated (based on \Str\
photometry) by Bond \& Perry (1971).

\section{The Standard Star Catalog}

Our final catalog of standard stars for \textit{uBVI} photometry is presented in
Table~4.  We list all 103 stars for which we have five or more observations. The
table gives the $u$ magnitudes and $u-B$ colors, along with the mean errors of
both quantities (derived from the internal scatter) and the number of $u$-band
observations that were averaged.  The $V$ magnitudes and $B-V$ and $V-I$ colors
of these stars are tabulated by L92.

The median mean error of our $u$ magnitudes is 0.006~mag.  Of the 103 standards,
97 have mean errors in $u$ of less than 0.020~mag, and of these 69 have errors
of less than 0.010~mag. Therefore, our catalog should easily provide
calibrations to better than 1\% accuracy.

In Figure~2 we plot the mean errors in the $u$ magnitude vs.\ the magnitude. As
expected, the errors increase with magnitude. If systematic errors remained in
the system, we would expect a flattened distribution. We compared the mean
errors with those output by DAOPHOT, and find that the mean errors are about
75\% larger on average than expected from photon statistics.  The larger
discrepancies are actually for the brighter stars, for which the formal errors
are almost entirely from photon statistics and are therefore small.  This
suggests that there is a small component of systematic error in our
magnitudes, arising from such issues as flat-fielding, inadequately modeled
red-leak corrections and photometric transformations, and short-term variations
in atmospheric extinction. In addition, some of the stars may be low-amplitude
variables.  Nevertheless, as noted above, these errors are small for most of our
standard stars, especially those with $u<16$.

Figure~3 shows a $u-B$ vs.\ $B-V$ diagram for the standard stars listed in
Table~4. The black symbols represent the standards from fields with galactic
latitudes $|b| \ge 30^\circ$; red symbols mark those in the three
low-latitude fields SA~98, Ru~149, and SA~110, which are likely to be reddened,
along with those in the SA~95 region, which is overlain with H$\alpha$ emission
and is thus also likely to be reddened. The figure confirms that these stars
are indeed systematically more reddened than the high-latitude stars.

\section{Sample Transformations}

In this section we present a check of the precision of our \textit{uBVI} standard-star
magnitudes, and give an example of calibration of data obtained with telescopes
other than the CTIO and KPNO 0.9-m reflectors used to establish the standard
stars. 

The data to be transformed are from a 1995 October CTIO 1.5-m run, and  a 1997
October KPNO 4-m run.  The CTIO 1.5-m observations employed the identical Tek3
CCD and $u$ filter used for the CTIO 0.9-meter observing runs, but different
$B$, $V$, and $I$ filters.  The KPNO 4-m observations, as compared with the KPNO
0.9-m, used a different CCD (T2KB), a different $u$ filter (but of the same
prescription), and different $B$, $V$, and $I$ filters.

Our intention in constructing this system is that observers should be able to
calibrate their observations to the \textit{uBVI} system without having to apply
red-leak corrections and nonlinear extinction terms. Being able to apply these
corrections would require detailed knowledge of the wavelength-dependent
throughput of the telescope system being used, and such measurements are
typically unavailable without significant effort (see also \S 5.3 of Paper~I
for further discussion). To test whether such a calibration will be
possible, we made the transformations to our standard system using only linear
and quadratic terms in color, and  linear and color terms in airmass; thus, we
are  allowing these terms to absorb the effects of the red leak and the small
quadratic extinction term.  The transformation equation used is the same as
eq.~12 of Paper~I, except that $k_2$ is set to zero.  We therefore have the
following equation:
\begin{equation}
u_{\rm instr} = u_{\rm std} + [a + b (B-V)] X + c + d (B-V) + e (B-V)^2 \, .
\end{equation}

We solved this equation using the matrix inversion code and techniques described
earlier (\S3.2).   Figure 4 shows the run of standard-star residuals for both
transformed observing runs.  The rms residuals for the $u$ magnitudes for both
runs are less than 1\%, demonstrating excellent photometric consistency. 
Moreover, the $u$ residuals are roughly comparable in size to the residuals of
our \textit{BVI} transformations for the same standard stars. Figure~4 does suggest a
slight downturn at the faintest magnitudes.  We have observed this trend in the
\textit{BVI} calibrations as well.  It is probably  the result of poor curve-of-growth
extrapolation for the faintest stars (which also have large error bars from the
photon statistics).  We typically remove these faint stars from the
calibrations, which still leaves a range of six magnitudes among the standards
that are retained.  

Our solutions for our  entire data archive of 0.9-, 1.5- and 4-m runs at both
CTIO and KPNO (with none of the data pre-corrected for red leak or extinction)
produced transformations of similar or better quality.  A typical transformation
uses two color terms: linear color (i.e., $d$), and one of  either the quadratic
color coefficient ($e$) or the color term in extinction ($b$). Typical values of
these coefficients are around $\pm$$0.05$~mag.

This exercise confirms that the effects of the red leak and non-linear
extinction can be corrected for fully by using linear and quadratic color terms,
as in eq.~8.  

\section{Conclusion}

We have defined a CCD-based photometric system that combines the Thuan-Gunn 
$u$ filter with the standard Johnson-Kron-Cousins \textit{BVI} filters. In this
paper we provide a catalog of $u$ magnitudes for 103 standard stars that can be
used (along with the \textit{BVI} magnitudes given by L92) to calibrate observations
taken through such filters.  We have shown that our standard system can be
reproduced using transformations of data from various telescope/detector
systems to an accuracy of better than 1\%.

Future papers will apply the \textit{uBVI} system to several programs involving
searches for luminous stars in young and old stellar systems, and
investigations of properties of the horizontal branch in globular clusters. We
hope this paper will encourage other observers to use the \textit{uBVI} system for
their own projects.

\begin{acknowledgments}
In addition to those thanked in Paper~I, we are grateful to C.~Palma for use of
his calibration code and for help with adapting that code for the purposes of
this study.  We also thank C.~Onken for helpful discussions.  This work was
supported in part by the NASA UV, Visible, and Gravitational Astrophysics
Research and Analysis Program through grants NAG5-3912 and NAG5-6821.
\end{acknowledgments}

\begin{figure}[h]
\begin{center}
\includegraphics[width=5.5in]{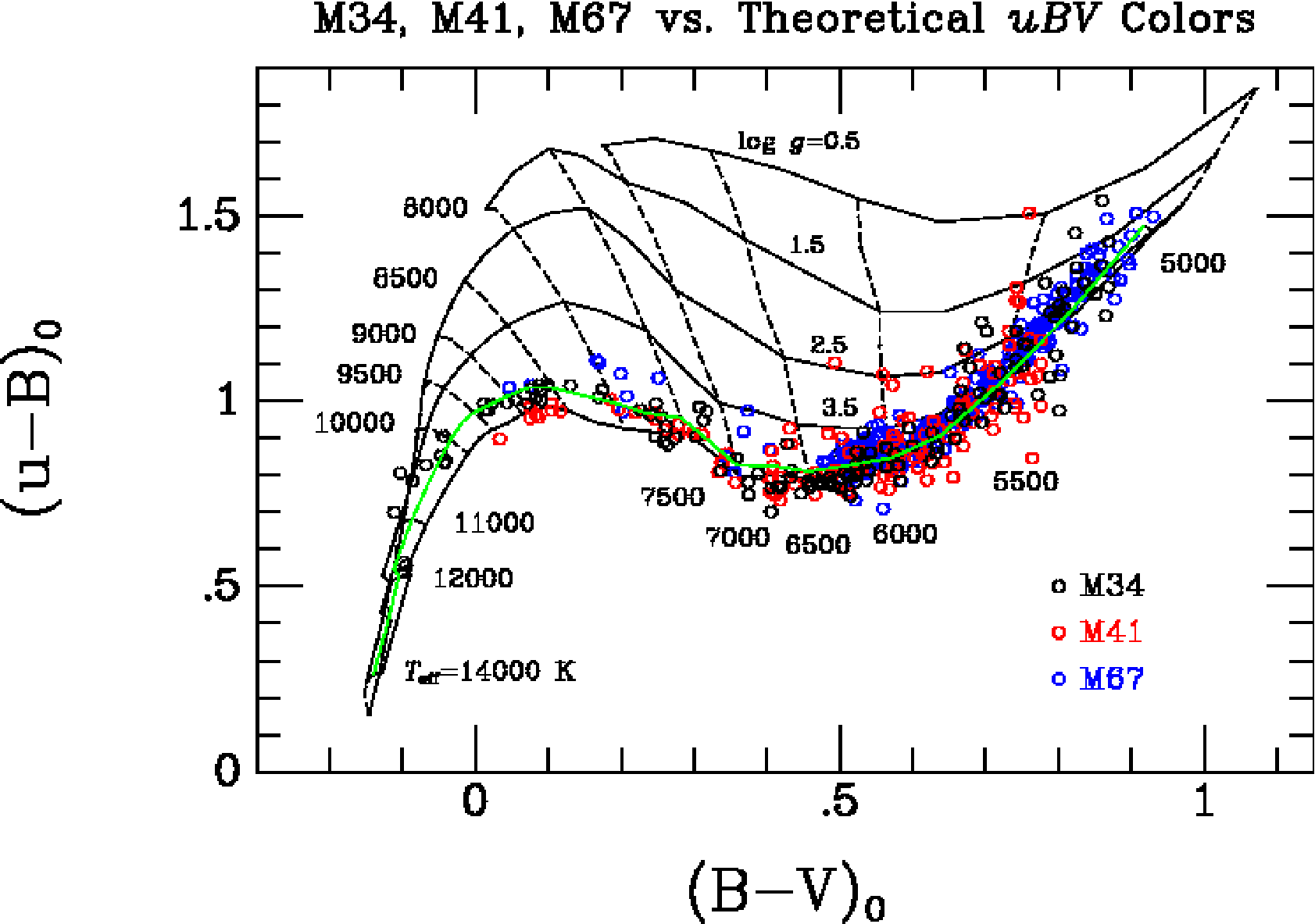}
\end{center}
\figcaption{Dereddened $(u-B)_0$ vs.\ $(B-V)_0$ diagram for observations of
three open clusters, M34 (black open circles), M41 (red open circles), and M67
(blue open circles). Adopted reddening values are given in Table~3. The grid of
theoretical colors for $\feh=0$ from Paper~I is superposed in black, and the
ZAMS relation from Paper~I is shown as a green line. The zero-point of our $u$
magnitude system was adjusted so that the main sequences of these three
clusters would agree in the mean with the theoretical relation, whose
zero-point is set such that Vega would have $u-B=1.0$.}
\end{figure}

\begin{figure}
\begin{center}
\includegraphics[width=5in]{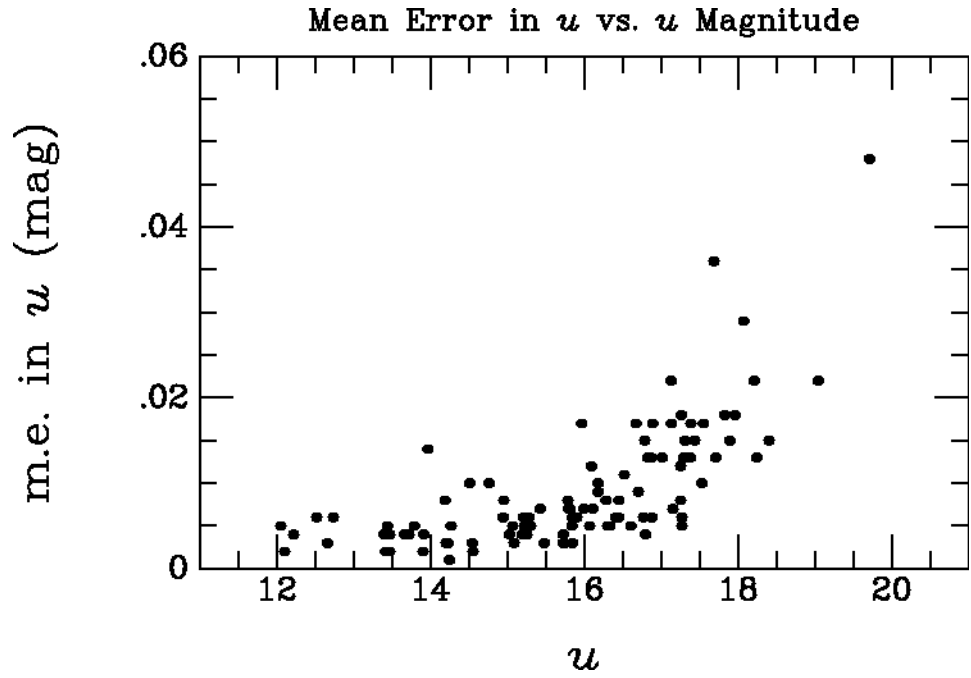}
\end{center}
\figcaption{The run of standard errors of the mean (m.e.)\ against $u$
magnitude for our \textit{uBVI} standard stars. Some of the high outliers may be
low-amplitude variables.}
\end{figure}

\begin{figure}
\begin{center}
\includegraphics[width=5.5in]{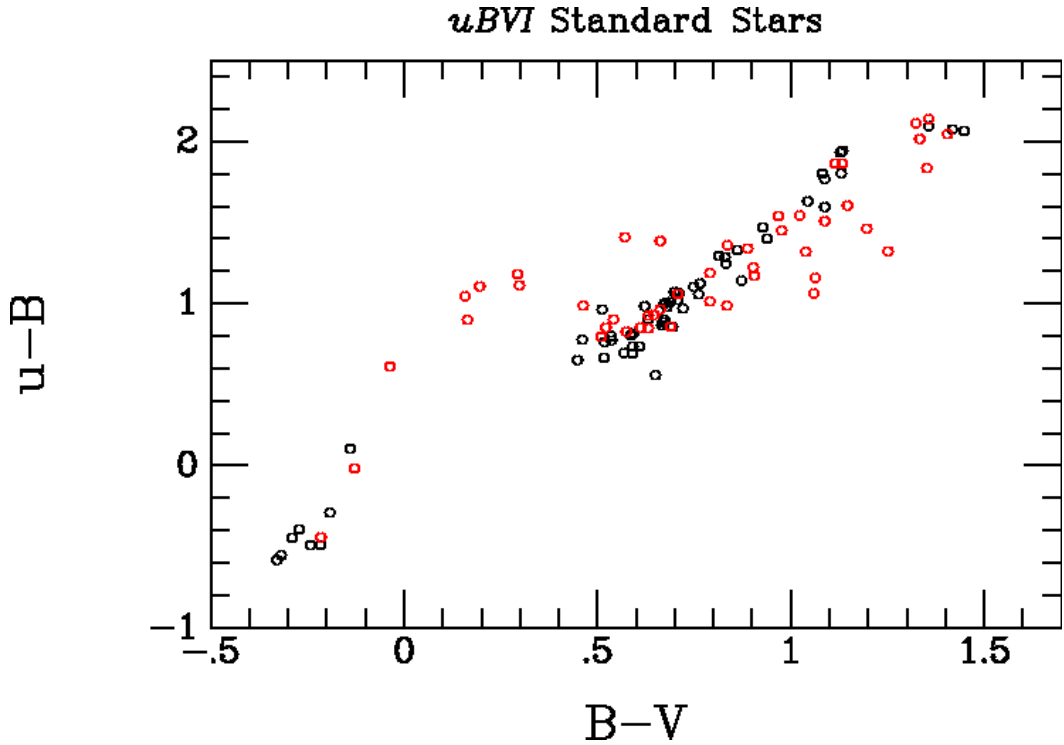}
\end{center}
\figcaption{$u-B$ vs.\ $B-V$ diagram for the standard stars listed in Table~4.
$B-V$ colors are taken from L92. Black open circles mark standards at high
galactic latitudes ($|b| \ge 30^\circ$). Red open circles mark those in three
low-latitude fields (SA~98, Ru~149, and SA~110) plus those in the reddened
SA~95 region; note that these stars are systematically more reddened than the
high-latitude stars.}
\end{figure}

\begin{figure}
\begin{center}
\includegraphics[width=6.25in]{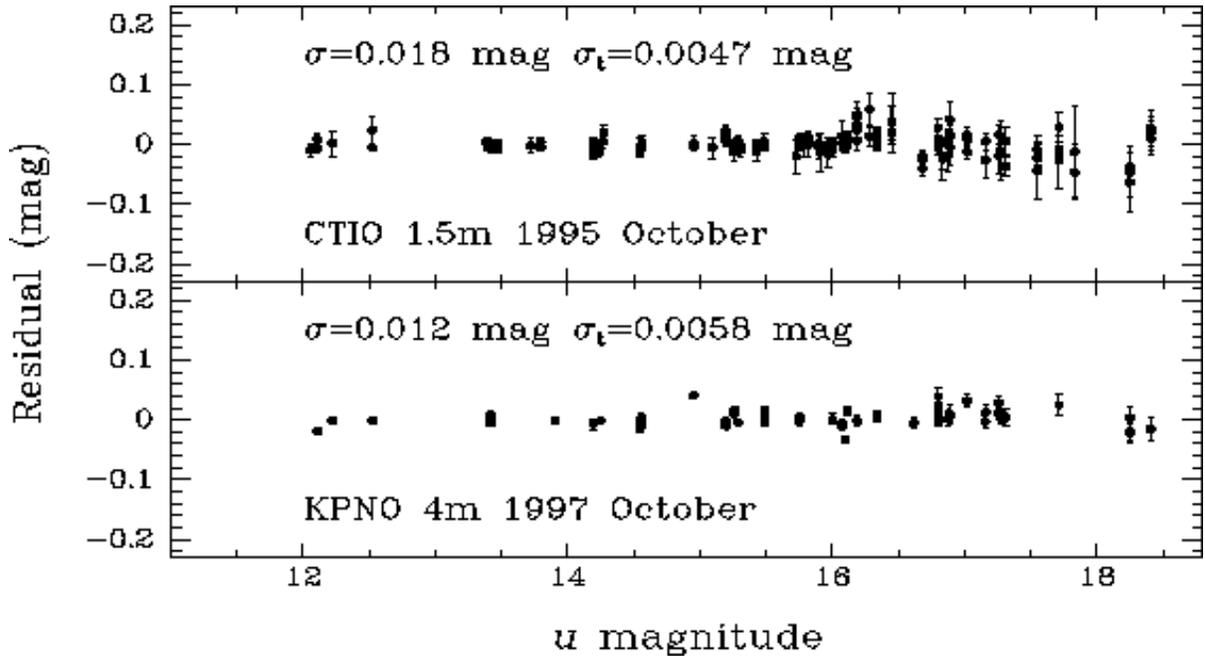}
\end{center}
\figcaption{Transformation errors for two independent observing runs.  Plotted
are the residuals of the fits against $u$ magnitudes of the standard stars. 
Given in the figure are both the unweighted scatter in the transformation
residuals ($\sigma$), and the uncertainty of the transformation from the
diagonal elements of the transformation matrix ($\sigma_t$) as  defined in
Harris, Fitzgerald, \& Reed (1981).}
\end{figure}

\begin{deluxetable}{lccc}
\tablewidth{0 pt}
\tablecaption{Standard Fields for \textit{uBVI} Photometry}
\tablehead{
\colhead{Field Name} &
\colhead{RA (J2000)} &
\colhead{Dec (J2000)} &
\colhead{Galactic latitude, $b$ ($^\circ$)} }
\startdata
SA 92$-$249     & 00:54:34 &  $+$00:41:05 & $-62$ \\
PG 0231+051   & 02:33:36 &  $+$05:19:00 & $-49$ \\
SA 95$-$112     & 03:53:40 &  $-$00:01:13 & $-39$ \\
SA 98$-$650     & 06:52:05 &  $-$00:19:40 & \phn$+0$ \\
Ru 149        & 07:24:15 &  $-$00:32:55 & \phn$+7$ \\
PG 0918+029   & 09:21:32 &  $+$02:47:00 & $+34$ \\
PG 1047+003   & 10:50:09 &  $-$00:01:08 & $+50$ \\
SA 104$-$334    & 12:42:21 &  $-$00:40:28 & $+62$ \\
PG 1323$-$086 & 13:25:52 &  $-$08:50:15 & $+53$ \\
PG 1633+099   & 16:35:32 &  $+$09:47:04 & $+35$ \\
SA 110$-$362    & 18:42:48 &  $+$00:06:26 & \phn$+2$ \\
Mark A        & 20:43:59 &  $-$10:46:42 & $-30$ \\
PG 2213$-$006 & 22:16:23 &  $-$00:21:45 & $-44$ \\
GD 246        & 23:12:20 &  $+$10:47:02 & $-45$ \\
\enddata
\end{deluxetable} 

\begin{deluxetable}{llcccc}
\tablewidth{0 pt}
\tablecaption{\textit{uBVI} 0.9-m Observing Runs}
\tablehead{
\colhead{Civil Dates} &
\colhead{Telescope+Detector} &
\colhead{$N_{\rm nights}$} &
\colhead{$N_{u\ \rm frames}$} &
\colhead{$N_{u\ \rm std\ stars}$} &
\colhead{$N_{u\ \rm measures}$}}
\startdata
1996 September 18--25  & KPNO 0.9m+T2KA & 4 & 16 & 38 & 96 \\ 
1997 May 7--10         & KPNO 0.9m+T2KA & 2 & 12 & 31 & 55 \\ 
1997 May 30--June 2    & CTIO 0.9m+Tek3 & 2 & 11 & 21 & 59 \\ 
1997 August 3--11      & CTIO 0.9m+Tek3 & 7 & 29 & 45 & \llap{1}62 \\ 
1997 September 17--23  & KPNO 0.9m+T2KA & 4 & 20 & 33 & 94 \\ 
1997 November 6--11    & CTIO 0.9m+Tek3 & 5 & 25 & 62 & \llap{1}51 \\ 
1998 March 17--23      & KPNO 0.9m+T2KA & 2 & 11 & 34 & 71 \\ 
1998 April 15--22      & CTIO 0.9m+Tek3 & 4 & 23 & 43 & \llap{1}10 \\ 
1998 August 18--27     & CTIO 0.9m+Tek3 & 8 & 42 & 71 & \llap{2}24 \\
1999 January 21--22    & KPNO 0.9m+T2KA & 1 & 10 & 42 & 73 \\
1999 March 12--16      & KPNO 0.9m+T2KA & 3 & 16 & 57 & \llap{1}32 \\
1999 June 10--15       & CTIO 0.9m+Tek3 & 2 & 11 & 22 & 49 \\ 
1999 August 24--27     & CTIO 0.9m+Tek3 & 3 & 14 & 32 & 73 \\ 
2001 March 24--28      & CTIO 0.9m+Tek3 & 3 & 16 & 26 & 67 \\ 
2001 November 9--13    & CTIO 0.9m+Tek3 & 4 & 15 & 50 & \llap{1}33 \\ 
\enddata
\end{deluxetable}

\begin{deluxetable}{lcl}
\tablewidth{0 pt}
\tablecaption{Open Clusters Used to Set the Zero Point}
\tablehead{
\colhead{Cluster} &
\colhead{$E(B-V)$} &
\colhead{Reference}}
\startdata
M34 (NGC 1039) & 0.07 & Jones \& Prosser 1996 \\
M41 (NGC 2287) & 0.03 & Harris et al.\ 1993 \\
M67 (NGC 2682) & 0.04 & Twarog et al.\ 1997 \\
\enddata
\end{deluxetable}

\begin{deluxetable}{lcrccr}
\tablewidth{0 pt}
\tablecaption{Catalog of \textit{uBVI} Standard Stars}
\tablehead{
\colhead{Star} &
\colhead{$u$} &
\colhead{$u-B$} &
\colhead{m.e.($u$)} &
\colhead{m.e.($u-B$)} &
\colhead{$n$}}
\startdata
SA 92$-$245    &   17.309 &    2.073 &    0.015 &   0.018 &   17\\
SA 92$-$248    &   18.405 &    1.931 &    0.015 &   0.034 &   18\\
SA 92$-$249    &   16.094 &    1.069 &    0.012 &   0.016 &   17\\
SA 92$-$250    &   15.284 &    1.292 &    0.006 &   0.008 &   17\\
SA 92$-$252    &   16.113 &    0.664 &    0.007 &   0.009 &   18\\
SA 92$-$253    &   17.018 &    1.802 &    0.013 &   0.015 &   17\\
SA 92$-$330    &   16.336 &    0.695 &    0.005 &   0.033 &   18\\
SA 92$-$335    &   14.194 &    0.999 &    0.008 &   0.009 &   18\\
SA 92$-$339    &   16.680 &    0.652 &    0.017 &   0.022 &   15\\
PG 0231+051  &   15.194 &   $-$0.582 &    0.004 &   0.012 &   27\\
PG 0231+051A &   14.543 &    1.061 &    0.003 &   0.004 &   26\\
PG 0231+051B &   18.247 &    2.064 &    0.013 &   0.015 &   27\\
PG 0231+051C &   15.257 &    0.884 &    0.004 &   0.009 &   28\\
PG 0231+051D &   16.884 &    1.769 &    0.006 &   0.010 &   28\\
PG 0231+051E &   15.486 &    1.005 &    0.003 &   0.007 &   27\\
SA 95$-$41     &   16.186 &    1.223 &    0.009 &   0.009 &   19\\
SA 95$-$42     &   14.949 &   $-$0.442 &    0.006 &   0.011 &   18\\
SA 95$-$43     &   12.108 &    0.795 &    0.002 &   0.004 &    5\\
SA 95$-$97     &   16.895 &    1.171 &    0.017 &   0.028 &    5\\
SA 95$-$100    &   17.437 &    1.013 &    0.015 &   0.085 &    5\\
SA 95$-$105    &   16.001 &    1.451 &    0.007 &   0.007 &   19\\
SA 95$-$107    &   19.711 &    2.112 &    0.048 &   0.117 &    7\\
SA 95$-$106    &   17.710 &    1.322 &    0.013 &   0.063 &   18\\
SA 95$-$112    &   17.549 &    1.385 &    0.017 &   0.017 &   20\\
SA 95$-$115    &   16.874 &    1.358 &    0.013 &   0.013 &   13\\
SA 98$-$590    &   17.830 &    1.836 &    0.018 &   0.024 &    5\\
SA 98$-$614    &   17.897 &    1.160 &    0.015 &   0.065 &    5\\
SA 98$-$624    &   15.788 &    1.186 &    0.008 &   0.029 &    6\\
SA 98$-$626    &   18.209 &    2.045 &    0.022 &   0.023 &    6\\
SA 98$-$627    &   16.448 &    0.859 &    0.008 &   0.020 &    6\\
SA 98$-$634    &   16.184 &    0.929 &    0.010 &   0.012 &    6\\
SA 98$-$642    &   17.269 &    1.408 &    0.018 &   0.052 &    6\\
SA 98$-$646    &   17.962 &    1.063 &    0.018 &   0.018 &    5\\
SA 98$-$650    &   13.474 &    1.046 &    0.002 &   0.003 &    6\\
SA 98$-$652    &   16.281 &    0.853 &    0.008 &   0.033 &    6\\
SA 98$-$666    &   13.795 &    0.899 &    0.005 &   0.007 &    6\\
SA 98$-$671    &   15.893 &    1.540 &    0.006 &   0.009 &    5\\
SA 98$-$670    &   15.425 &    2.139 &    0.007 &   0.007 &    6\\
SA 98$-$676    &   15.819 &    1.605 &    0.007 &   0.009 &    6\\
SA 98$-$682    &   15.306 &    0.925 &    0.005 &   0.008 &    6\\
SA 98$-$685    &   13.402 &    0.985 &    0.004 &   0.006 &    6\\
SA 98$-$688    &   14.230 &    1.183 &    0.003 &   0.005 &    6\\
SA 98$-$1002   &   15.966 &    0.824 &    0.017 &   0.019 &    5\\
SA 98$-$1082   &   16.830 &    0.985 &    0.013 &   0.020 &    5\\
Ru 149       &   13.721 &    $-$0.016 &    0.004 &   0.006 &   14\\
Ru 149A      &   15.906 &    1.113 &    0.006 &   0.010 &   14\\
Ru 149B      &   14.268 &    0.964 &    0.005 &   0.006 &   14\\
Ru 149C      &   15.727 &    1.107 &    0.003 &   0.008 &   14\\
Ru 149D      &   12.054 &    0.611 &    0.005 &   0.006 &   14\\
Ru 149E      &   15.091 &    0.851 &    0.003 &   0.008 &   14\\
Ru 149F      &   16.449 &    1.863 &    0.006 &   0.010 &   14\\
Ru 149G      &   14.272 &    0.902 &    0.005 &   0.007 &   14\\
PG 0918+029  &   12.663 &   $-$0.393 &    0.003 &   0.005 &    6\\
PG 0918+029A &   15.799 &    0.773 &    0.007 &   0.010 &    6\\
PG 0918+029B &   15.852 &    1.124 &    0.006 &   0.010 &    6\\
PG 0918+029C &   15.070 &    0.902 &    0.005 &   0.006 &    6\\
PG 0918+029D &   14.947 &    1.631 &    0.006 &   0.007 &    6\\
PG 1047+003  &   12.739 &   $-$0.445 &    0.006 &   0.008 &   18\\
PG 1047+003A &   15.210 &    1.010 &    0.006 &   0.009 &   18\\
PG 1047+003B &   16.409 &    0.979 &    0.006 &   0.012 &   17\\
SA 104$-$237   &   18.079 &    1.596 &    0.029 &   0.029 &    6\\
SA 104$-$239   &   17.388 &    2.096 &    0.017 &   0.020 &   11\\
SA 104$-$244   &   17.292 &    0.691 &    0.013 &   0.017 &   10\\
SA 104$-$325   &   17.131 &    0.856 &    0.017 &   0.049 &   11\\
SA 104$-$330   &   16.707 &    0.817 &    0.009 &   0.031 &   11\\
SA 104$-$334   &   14.765 &    0.763 &    0.010 &   0.012 &   11\\
SA 104$-$336   &   16.521 &    1.287 &    0.011 &   0.015 &   11\\
SA 104$-$338   &   17.386 &    0.736 &    0.013 &   0.027 &   11\\
SA 104$-$339   &   17.535 &    1.244 &    0.010 &   0.010 &   10\\
SA 104$-$423   &   17.134 &    0.902 &    0.022 &   0.051 &    6\\
SA 104$-$444   &   14.953 &    0.964 &    0.008 &   0.013 &    6\\
SA 104$-$456   &   13.968 &    0.984 &    0.014 &   0.014 &    6\\
SA 104$-$L2    &   17.256 &    0.558 &    0.012 &   0.035 &   11\\
PG 1323$-$086  &   13.444 &    0.103 &    0.005 &   0.006 &   19\\
PG 1323$-$086C &   15.730 &    1.020 &    0.004 &   0.006 &   19\\
PG 1323$-$086B &   15.223 &    1.056 &    0.005 &   0.006 &   19\\
PG 1323$-$086D &   13.474 &    0.807 &    0.004 &   0.005 &   18\\
PG 1633+099  &   13.914 &   $-$0.291 &    0.004 &   0.005 &   26\\
PG 1633+099A &   17.270 &    1.141 &    0.006 &   0.008 &   24\\
PG 1633+099B &   15.850 &    1.800 &    0.003 &   0.004 &   25\\
PG 1633+099C &   16.307 &    1.944 &    0.005 &   0.006 &   26\\
PG 1633+099D &   15.025 &    0.799 &    0.004 &   0.005 &   25\\
SA 110$-$266   &   14.248 &    1.341 &    0.001 &   0.003 &   26\\
SA 110$-$290   &   13.664 &    1.058 &    0.004 &   0.005 &    6\\
SA 110$-$349   &   17.691 &    1.508 &    0.036 &   0.036 &    6\\
SA 110$-$355   &   14.512 &    1.545 &    0.010 &   0.010 &    6\\
SA 110$-$358   &   16.787 &    1.318 &    0.015 &   0.015 &    6\\
SA 110$-$360   &   17.275 &    1.460 &    0.005 &   0.022 &   25\\
SA 110$-$361   &   13.904 &    0.847 &    0.002 &   0.004 &   26\\
SA 110$-$362   &   19.042 &    2.017 &    0.022 &   0.022 &   23\\
SA 110$-$364   &   16.613 &    1.865 &    0.005 &   0.009 &   26\\
Mark A       &   12.524 &   $-$0.492 &    0.006 &   0.006 &   23\\
Mark A1      &   17.256 &    0.736 &    0.008 &   0.012 &   23\\
Mark A2      &   16.072 &    0.866 &    0.005 &   0.007 &   23\\
Mark A3      &   17.156 &    1.400 &    0.007 &   0.008 &   23\\
PG 2213$-$006  &   13.418 &   $-$0.489 &    0.002 &   0.004 &   35\\
PG 2213$-$006A &   15.751 &    0.900 &    0.003 &   0.007 &   35\\
PG 2213$-$006B &   14.558 &    1.103 &    0.002 &   0.003 &   35\\
PG 2213$-$006C &   16.798 &    0.968 &    0.004 &   0.008 &   34\\
GD 246       &   12.221 &   $-$0.555 &    0.004 &   0.004 &   13\\
GD 246A      &   14.203 &    0.777 &    0.003 &   0.007 &   13\\
GD 246B      &   16.771 &    1.472 &    0.006 &   0.009 &   13\\
GD 246C      &   15.842 &    1.332 &    0.005 &   0.008 &   13\\
\enddata
\end{deluxetable}

\end{document}